\documentstyle[prd,aps,epsfig,amssymb]{revtex}
\let\jnfont=\rm
\def\NPB#1,{{\jnfont  Nucl.\ Phys.\ B }{\bf #1},}
\def\PLB#1,{{\jnfont Phys.\ Lett.\ B }{\bf #1},}
\def\EPJC#1,{{\jnfont Euro.\ Phys.\ J.\ C }{\bf #1},}
\def\PRD#1,{{\jnfont \em Phys.\ Rev.\ D }{\bf #1},}
\def\PRL#1,{{\jnfont Phys.\ Rev.\ Lett.\ }{\bf #1},}
\def\MPLA#1,{{\jnfont Mod.\ Phys.\ Lett.\ A }{\bf #1},}
\def\JPG#1,{{\jnfont J.\ Phys.\ G}{\bf #1},}
\def\CTP#1,{{\jnfont Commun.\ Theor.\ Phys.\ }{\bf #1},}

\def\p_slash{\not{\hbox{\kern-2.1pt $p$}}}
\def\k_slash{\not{\hbox{\kern-2.1pt $k$}}}
\def\E_slash{\not{\hbox{\kern-2.1pt $E$}}}

\begin{document}
\draft
\preprint{}

\title{Probing Topcolor-Assisted Technicolor from Like-sign Top Pair Production at LHC}
\author{Junjie Cao $^{a,b}$, Guoli Liu $^b$,  Jin Min Yang $^{b}$ \\\ }
\address{$^a$ Department of Physics, Henan Normal University, Henan 453002, China}
\address{$^b$ Institute of Theoretical Physics, Academia Sinica, Beijing 100080, China}

\date{\today}
\maketitle

\begin{abstract}
The topcolor-assisted technicolor (TC2) theory predicts tree-level
flavor-changing neutral-current (FCNC) top quark Yukawa couplings
with top-pions. Such FCNC interactions will induce like-sign top
quark pair productions at CERN Large Hadron Collider (LHC). While
these rare productions are far below the observable level in the
Standard Model and other popular new physics models such as the
Minimal Supersymmetric Model, we find that in a sound part of
parameter space the TC2 model can enhance the production cross
sections to several tens of fb and thus may be observable at the
LHC due to rather low backgrounds. Searching for these productions
at the LHC will serve as an excellent probe for the TC2 model.

\end {abstract}

\section{Introduction}
Top quark physics \cite{review} will be intensively studied in the
coming years. The Fermilab Tevatron Collider and the CERN Large
Hadron Collider (LHC) will copiously produce top quarks and allow
to scrutinize top quark properties.  Any new physics related to
top quark will be uncovered or stringently constrained
\cite{sensitive}. One striking property of top quark in the
Standard Model (SM) is its extremely weak flavor-changing
neutral-current (FCNC) interactions due to the GIM mechanism: they
are absent at tree-level and highly suppressed at loop-level
\cite{tcvh_sm}. By contrast, the extensions of the SM often
inevitably predict much larger FCNC interactions for top quark.
Therefore, the study of top quark FCNC processes will serve as a
sensitive test of the SM and a powerful probe of new physics.

In the extensions of the SM, the top quark FCNC interactions may
be enhanced through two ways. One is that at loop-level the GIM
machanism does not work so well as in the SM since new particles
enter the loops to mediate top quark FCNC transitions. The other
is that some models natually predict tree-level top quark FCNC
Yukawa couplings with scalar fields, which is in contrast with the
SM where the generation of fermion masses is realized by simply
introducing Yukawa couplings with only one Higgs doublet and, as a
result, the Yukawa couplings can be diagonalized simultaneously
with the fermion mass matrices. The enhanced top quark FCNC
interactions will lead to various possibly observable FCNC
processes at colliders, such as the FCNC decays
\cite{tcv1,tcv2,tcv3} and the top-charm associated
productions\cite{cao1,yue}. In addition, they can also induce the
like-sign top pair productions at the LHC. Unlike top-charm
associated productions, these like-sign top pair productions are
free from huge QCD background $W+jets$ and also from $t \bar{t}$
background \cite{tc-background}. Due to rather low backgrounds
\cite{tt-background}, such productions will be an excellent probe
for top quark FCNC interactions \cite{Like-sign}.

In this article, we study the possibility of using the like-sign top pair productions at the LHC
to probe the topcolor-assisted technicolor (TC2) theory \cite{tc2-Hill,tc2-Lane}.
This theory, which combines the fancy idea of technicolor\cite{techni} with top quark
condensation\cite{top-condensation}, has not yet been excluded by experiments and remains
a typical candidate of new physics in the direction of dynamical electroweak symmetry breaking (EWSB).
A remarkable feature of this theory is that it predicts tree-level FCNC Yukawa interactions
for top quark since top quark is singled out for condensation to generate the main part
of its mass \cite{FCNH,top-Higgs}.
Such tree-level FCNC interactions are likely
to induce sizable like-sign top pair productions at the LHC.
Since these rare productions are far below the observable level in the SM
and other popular new physics models like supersymmetry (we will discuss and estimate later),
the observation/unobserevation of these productions will strongly favor/disfavor
the TC2 theory.

This paper is organized as follows. In Section II, we first
briefly introduce the TC2 theory and then recapitulate the current theoretical and
experimental constraints on the parameters of this theory. In Section
III, we calculate various like-sign top pair productions at the LHC
induced by top quark FCNC interactions in the TC2 theory and discuss
their observability. We also discuss about the predictions of other popular new
physics models. Finally in Section IV we give the conclusion.

\section{Topcolor-assisted technicolor}

The TC2 theroy \cite{tc2-Hill,tc2-Lane} introduces two strongly
interacting sectors, with one sector (topcolor interaction)
generating the large top quark mass and partially contributing to
EWSB while the other sector
(technicolor interaction) responsible for the bulk of EWSB and the
generation of light fermion masses.  At the EWSB scale, it
predicts the existence of two groups of composite scalars from
topcolor and technicolor condensations, respectively
\cite{tc2-Hill,tc2-Lane,top-condensation}. In the linear
realization,  the scalars of our interest can be arranged into
two $SU(2)$ doublets, namely  $\Phi_{top}$ and $\Phi_{TC}$
\cite{top-condensation,2hd,Rainwater}, which are analogous to the
Higgs fields in a special two-Higgs-doublet model \cite{special}.
The doublet $\Phi_{top}$ from topcolor condensation couples only
to the third-generation quarks. Its main task is to generate the
large top quark mass. It can also generate a sound part of bottom
quark mass indirectly via instanton effect\cite{tc2-Hill}. Since a
small value of the top-pion decay constant $F_t $ (the vev of the
doublet $\Phi_{top}$) is theoretically favored (see below), this
doublet must couple strongly to top quark in order to generate the
expected top quark mass. The other doublet $\Phi_{TC}$, which is
technicolor condensate,  is mainly responsible for EWSB and light
fermion masses. It also contributes a small portion to the 
third-generation quark masses. Because its vev $v_{TC}$ is generally
comparable with $v_W$, its Yukawa couplings with all fermions are
small. The low-energy effective Lagrangian can be written as
\cite{Rainwater}
\begin{eqnarray}
{\cal L}= | D_{\mu} \Phi_{TC} |^2 + | D_{\mu} \Phi_{top} |^2 -
\left ( \sum_{i,j=1}^3 \lambda_{i j}^U \bar{Q}_{L i} \Phi_{TC}
U_{R j} + \sum_{i,j=1}^3 \lambda_{i j}^D \bar{Q}_{L i}
\tilde{\Phi}_{TC} D_{R j} + Y_t \bar{\Psi}_L \Phi_{top} t_R + h.c.
\right ) + \cdots
\end{eqnarray}
where $ D_{\mu} = \partial_{\mu}+ ig' \frac{Y}{2} B_{\mu} + ig
\frac{\tau_i}{2} W_{\mu}^i$, $Q_{L i}$ denotes the left-handed
quark doublet, $U_{R j}$ and $ D_{R j}$ are right-handed quarks,
$\Psi_L $ is the left-handed top-bottom doublet,
$\tilde{\Phi}_{TC} $ is the conjugate of $\Phi_{TC}$, and
$\lambda_{i j}^{U, D}$ and $Y_t $ are Yukawa coupling constants
satisfying $\lambda_{i j}^{U, D}\ll Y_t$. The two $SU(2)$ doublets
take the form
\begin{eqnarray}
\Phi_{TC}& =&\left ( \begin{array}{c} v_{TC} + ( H_{TC}^0 + i \Pi_{TC}^0 )/\sqrt{2}  \\
           \Pi_{TC}^- \end{array} \right ) , \\
\Phi_{top}& =& \left (\begin{array}{c} F_t + ( H_{top}^0 + i \Pi_{top}^0 )/\sqrt{2}  \\
           \Pi_{top}^- \end{array} \right ) .
\end{eqnarray}
We can rotate the two doublets into $\Phi_{1,2}$ such that
$<\Phi_1>=\sqrt{v_{TC}^2 + F_t^2}=v_w$ and $<\Phi_2> =0$
\begin{eqnarray}
\Phi_1 & = & (\cos \beta \Phi_{TC} + \sin \beta \Phi_{top} )=
            \left ( \begin{array}{c} v_{w} + ( H_1^0+ i G^0 )/\sqrt{2}  \\
             G^- \end{array} \right ) , \\
\Phi_2 & = & (- \sin \beta \Phi_{TC} + \cos \beta \Phi_{top} )
            =\left (
            \begin{array}{c} ( H_2^0 + i A^0 )/\sqrt{2} , \\
            H^- \end{array} \right ) ,
\end{eqnarray}
where $\tan \beta =F_t/v_{TC}$. Then the Lagrangian can be rewritten as
\begin{eqnarray}
{\cal L}&=&|D_{\mu} \Phi_{1}|^2 + |D_{\mu} \Phi_{2}|^2 -
\left ( \sum_{i,j=1}^3 \lambda_{i j}^{\prime U} \bar{Q}_{L i}
\Phi_{1} U_{R j} + \sum_{i,j=1}^3 \lambda_{i j}^D
\frac{\sqrt{v_w^2-F_t^2}}{v_w} \bar{Q}_{L i} \tilde{\Phi}_{1} D_{R
j}  \right. \nonumber   \\  & &  \left. - \sum_{i,j=1}^3
\lambda_{i j}^D \frac{F_t}{v_w} \bar{Q}_{L i} \tilde{\Phi}_{2}
D_{R j} - \sum_{i,j=1}^3 \lambda_{i j}^U \frac{F_t}{v_w}
\bar{Q}_{L i} \Phi_2 U_{R j}+ Y_t \frac{\sqrt{v_w^2-F_t^2}}{v_w}
\bar{\Psi}_L \Phi_{2} t_R + h.c. \right ) + \cdots
\label{Lagrangian}
\end{eqnarray}
where $\lambda_{i j}^{\prime U} =\lambda_{i j}^U \cos \beta + Y_t
\sin \beta \delta_{i 3} \delta_{j 3} $. In this new basis, $G^\pm$
and $G^0$ are Goldstone bosons while the pseudoscalar $A^0$,
the charged scalar $H^\pm$ and the CP-even scalars $H_{1,2}^0$ are
physical Higgs bosons.
It is obvious that $H_1^0$ plays the role of the "standard"
Higgs boson with flavor diagonal couplings and  $H_2^0$ decouples from
the SM vector bosons but has strong coupling only with top quark.
In our following analysis, we will adopt the same notations as in the literature,
i.e.,  using top-Higgs $h_t^0$, top-pions $\pi_t^{0,\pm}$ to denote $H_2^0$, $A^0$
and $H^\pm$, respectively.

In Eq.(\ref{Lagrangian}), the rotation of quarks into their mass
eigenstates will induce FCNC Yukawa interactions from the $\Phi_2$
couplings \footnote{ Just like the Higgs field in the SM, $\Phi_1$
terms give no FCNC couplings since they are diagonalized
simultaneously with the fermion mass matrices.}. Since $\lambda_{i
j}^{U, D}\ll Y_t $, the FCNC couplings from $\lambda_{i j}^U$ and
$\lambda_{i j}^D$ can be safely neglected.  Because $Y_t
=(1-\epsilon) m_t/F_t$ ($\epsilon $ denoting the fraction of
technicolor contribution to the top quark mass) is quite large
(about $2 \sim 3 $) and the mixing between $c_R$ and $t_R$ can be
natually as large as 30\% \cite{FCNH}, the FCNC coupling from the
$Y_t$ term may be sizable and thus may have significant
phenomenological consequence. The FCNC couplings from this term
are given by
\begin{eqnarray}
{\cal{L}}_{FCNC}& = &\frac{(1 - \epsilon ) m_{t}}{\sqrt{2}F_{t}}
     \frac{\sqrt{v_{w}^{2}-F_{t}^{2}}} {v_{w}} \left (
            i K_{UL}^{tt*}K_{UR}^{tt } \bar{t}_L t_{R} \pi_t^0
           + \sqrt{2}K_{UR}^{tt *} K_{DL}^{bb}\bar{t}_R b_{L} \pi_t^-
           + i K_{UL}^{tt *} K_{UR}^{tc} \bar{t}_L c_{R} \pi_t^0  \right . \nonumber \\
& & \left. + \sqrt{2} K_{UR}^{tc *} K_{DL}^{bb} \bar{c}_R b_{L} \pi_t^-
           + K_{UL}^{tt*} K_{UR}^{tt } \bar{t}_L t_{R} h_t^0
           + K_{UL}^{tt *} K_{UR}^{tc} \bar{t}_L c_{R} h_t^0 + h.c.  \right ) ,
\label{FCNH}
\end{eqnarray}
where $K_{UL}$, $K_{DL}$ and $K_{UR}$ are the rotation matrices that
transform the weak eigenstates of left-handed up-type, down-type and
right-handed up-type quarks to their mass eigenstates, respectively.
According to the analysis of \cite{FCNH}, their favored values are
given by
\begin{equation}
K_{UL}^{tt} \simeq K_{DL}^{bb} \simeq 1, \hspace{5mm}
K_{UR}^{tt}\simeq \frac{m_t^\prime}{m_t} = 1-\epsilon,
\hspace{5mm} K_{UR}^{tc}\leq \sqrt{1-(K_{UR}^{tt})^2}
=\sqrt{2\epsilon-\epsilon^{2}},
\label{FCSI}
\end{equation}
with $m_t^\prime$ denoting the topcolor contribution to the top
quark mass. In Eq.(\ref{FCNH}) we neglected the mixing
between up quark and top quark.

Now we recapitulate the theoretical and experimental constraints on the
relevent parameters.
\begin{itemize}
\item[{\rm (1)}] About the $\epsilon$ parameter. In the TC2 model,
$\epsilon $ parameterizes the portion of the extended-technicolor (ETC)
contribution to the top quark mass. The bare value of $\epsilon $ is generated at the
ETC scale, and subject to very large radiative enhancement from
the topcolor and $U(1)_{Y_1}$ by a factor of order $10$ when evolving
down to the weak scale \cite{tc2-Hill}. This $\epsilon$ can induce
a nonzero top-pion mass (proportional to $\sqrt{\epsilon} $) \cite{Hill}
and thus ameliorate the problem of having dangerously light scalars.
Numerical analysis shows that, with reasonable choice of other input parameters,
$\epsilon$ of order $10^{-2} \sim 10^{-1}$ may induce top-pions as massive as the top
quark \cite{tc2-Hill}. Indirect phenomenological constraints on $\epsilon $
come from low energy flavor-changing processes such as $b \to s \gamma$ \cite{b-sgamma}.
However, these constraints are very weak. From the theoretical point of view, $\epsilon $
with value from $0.01$ to $0.1$ is favored. Since a large $\epsilon$ can slightly suppress
the FCNC Yukawa couplings, we fix conservatively $\epsilon =0.1$ throughout this paper.

\item[{\rm (2)}] The parameter $K_{UR}^{tc}$ is upper bounded by the unitary relation
$K_{UR}^{tc} \leq \sqrt{1-(K_{UR}^{tt})^ 2}=\sqrt{2\epsilon -\epsilon^2}$.
For a $\epsilon $ value smaller than $0.1 $, this corresponds to $ K_{UR}^{tc} < 0.43$.
In our analysis, we will treat $K_{UR}^{tc}$ as a free parameter.

\item[{\rm (3)}] About the top-pion decay constant $F_t$,  the Pagels-Stokar formula \cite{Pagels}
gives an expression in terms of the number of quark color $N_c$, the top quark mass, and
the scale $\Lambda $ at which the condensation occurs:
\begin{eqnarray}
F_t^2= \frac{N_c}{16 \pi^2} m_t^2 \ln{\frac{\Lambda^2}{m_t^2}}.
\label{ft}
\end{eqnarray}
From this formula, one can infer that, if $t\bar{t} $ condensation is fully responsible for EWSB,
i.e. $F_t \simeq v_w \equiv v/\sqrt{2} = 174$ GeV, then $\Lambda $ is about
$10^{13} \sim 10^{14}$ GeV. Such a large value is less attractive since by the
original idea of technicolor \cite{techni}, one expects new physics scale should not be far
higher than the weak scale. On the other hand, if one believes that new physics exists at TeV scale,
i.e. $\Lambda \sim 1$ TeV, then $F_t \sim 50$ GeV, which means that $t \bar{t} $ condensation alone
cannot be wholly responsible for EWSB and to break electroweak symmetry needs the joint effort of
topcolor and other interactions like technicolor. By the way, Eq.(\ref{ft}) should be understood as
only a rough guide, and $F_t$ may in fact be somewhat lower or higher, say in the range $40 \sim 70$ GeV.
Allowing $F_t $ to vary over this range does not qualitatively change our conclusion, and, therefore,
we use the value $F_t =50$ GeV for illustration in our numerical analysis.

\item[{\rm (4)}] About the mass bounds for top-pions and top-Higgs.
On the theoretical side, some estimates have been done. The mass splitting between
the neutral top-pion and the charged top-pion should be small since it comes only
from the electroweak interactions \cite{mass-pion}. Ref.\cite{tc2-Hill} has estimated
the mass of top-pions using quark loop approximation and showed that $m_{\pi_t}$ is
allowed to be a few hundred GeV in a reasonable parameter space. Like Eq.(\ref{ft}),
such estimations can only be regarded as a rough guide and the precise values of top-pion
masses can be determined only by future experiments. The mass of the top-Higgs $h_t^0$
can be estimated in the Nambu-Jona-Lasinio (NJL) model in the large $N_{c}$ approximation
and is found to be about $2m_{t}$ \cite{top-condensation,top-Higgs}. This estimation
is also rather crude and the mass below the $\overline{t}t$ threshold is quite possible
in a variety of scenarios \cite{y15}. On the experimental side,  current experiments
have restricted the mass of the charged top-pion. For example, the absence of $t \to \pi_t^+b$
implies that $m_{\pi_t^+} > 165$ GeV \cite{t-bpion} and $R_b$ analysis yields
$m_{\pi_t^+}> 220$ GeV \cite{burdman,kuang}. For the neutral top-pion and top-Higgs,
the experimental restrictions on them are rather weak. (Of course, considering
theoretically that the mass splitting between the neutral and charged top-pions is small,
the $R_b$ bound on the charged top-pion mass should be applicable to the neutral top-pion
masses.)  The current bound on techni-pions \cite{datagroup} does not apply here since the
properties of top-pion are quite different from those of techni-pions. The direct search
for the neutral top-pion (top-Higgs) via $ p p ({\rm or}~p\bar p) \to t \bar{t} \pi_t^0 (h_t^0)$ with
$\pi_t^0 (h_t^0) \to b \bar{b} $ was proven to be hopeless at Tevatron for the top-pion
(top-Higgs) heavier than $135 $ GeV \cite{Rainwater}. The single production of $\pi_t^0 $
($h_t^0$ ) at Tevatron with $\pi_t^0 $ ($h_t^0$) mainly decaying to $t \bar{c} $ may shed
some light on detecting top-pion (top-Higgs)\cite{top-Higgs}, but the potential for the
detection is limited by the value of $K_{UR}^{tc}$ and the detailed background analysis
is absent now. Anyhow, these mass bounds will be greatly tightened  at the upcoming LHC
\cite{cao1,FCNH,Rainwater}. Combining the above theoretical and experimental bounds,
we in our discussion will assume
\begin{equation}
m_{h_t^0} > 135 ~{\rm GeV} \hspace{5mm}
m_{\pi_{t}^{0}}=m_{\pi_{t}^+}\equiv m_{\pi_t} > 220 ~{\rm GeV} .
\end{equation}
\end{itemize}
\begin{figure}[hbt]
\begin{center}
\epsfig{file=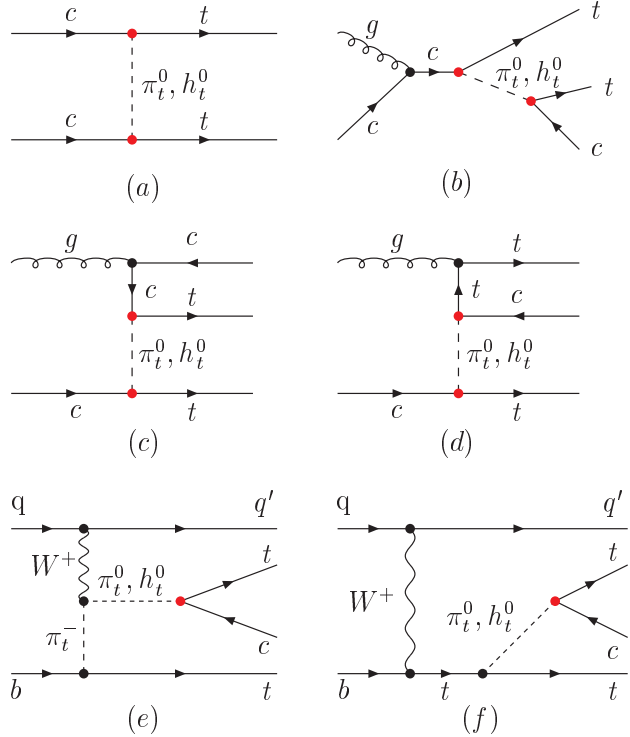,width=9cm}
\caption{Feynman diagrams for like-sign top pair productions induced by the FCNC Yukawa
         interactions in the TC2 model. }
\label{feynman}
\end{center}
\end{figure}

\section{Like-sign top pair productions at LHC}

Due to the existence of the top quark FCNC Yukawa interactions in Eq.(\ref{FCNH}),
the like-sign top pair productions can proceed through various parton processes
at the LHC, as shown in Fig.\ref{feynman}.
Since the signals of these processes as well as their corresponding backgrounds are
different, we will analysis these processes separately.
Throughout this paper, we take $m_t =178 $ GeV \cite{topmass}, $m_w=80.448 $
GeV \cite{datagroup}, $\alpha_s(m_z) = 0.118 $ and neglect bottom
quark mass as well as charm quark mass. We used CTEQ6L \cite{CTEQ} parton distribution
functions with scale $\mu =2m_t$.

\subsection{  $t t$ production at the LHC}

\begin{figure}[hbt]
\begin{center}
\epsfig{file=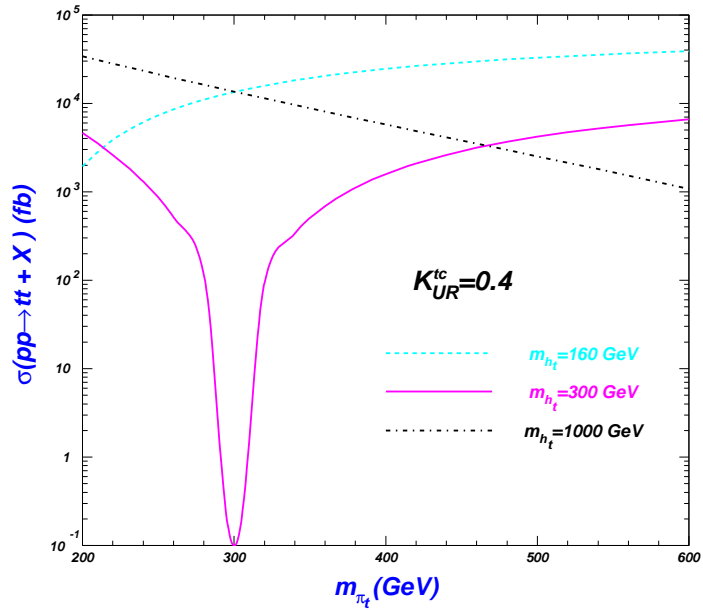,width=10cm}
\caption{Cross section of $ p p \to t t +X$ at the LHC as a function of $m_{\pi_t}$.}
\label{pptt}
\end{center}
\end{figure}

In the TC2 model, $ pp  \to t t + X$ proceeds through the patron process
$c c \to t t$ by exchanging a neutral top-pion or top-Higgs, as shown in
Fig.\ref{feynman} (a).
This process has two characters. One is that its cross section is proportional to
$\left ( K_{UR}^{tc} \right )^4 $ in all the parameter space, and thus
very sensitive to $K_{UR}^{tc}$.
The other is that the top-pion diagrams and the top-Higgs diagrams
interfer destructively and such destructive effect is significant
for degenerate top-pion and top-Higgs masses. This feature is illustrated
in Fig.\ref{pptt}  for three representative values of $m_{h_t}$.
For a light top-Higgs with $m_{h_t} =160$ GeV, the increase of the
cross section as top-pion becomes heavier is due to the weakening
cancellation effect. For a moderate top-Higgs with $ m_{h_t}=300 $ GeV,
the dip of the cross section as $m_{\pi_t}$ approaches $m_{h_t}$
is a direct reflection of the cancellation effect. For a heavy top-Higgs
$m_{h_t} =1000$ GeV, the top-Higgs contribution is strongly
suppressed relative to the top-pion contribution and the total
cross section is dominated by  the top-pion contribution. As a result,
the total cross section decreases monotonously as the top-pions get heavier,
showing the decoupling effects.

Note that in Fig.\ref{pptt} we fix $K_{UR}^{tc} =0.4$ and
the charge conjugate production $pp \to \bar t \bar t + X$ is
also taken into account.
The cross section for an arbitrary $K_{UR}^{tc}$ value can
be obatined by scaling the result of Fig.\ref{pptt} by a factor of
$ \left (K_{UR}^{tc}/0.4 \right )^4$. So one can infer that even for
$K_{UR}^{tc} =0.1$, the cross section can still reach the level of
several fb in a vast parameter space.
Now we discuss the observability of the production  $pp \to t t + X$
and its  charge conjugate production channel.
The semileptonical decay of both top (or anti-top) quarks give rise to
a signal of like-sign dilepton plus two b-jets, i.e., $\ell^\pm \ell^\pm + 2$ b-jets
($\ell=e,\mu$).
The major backgrounds are from the production of $t \bar{t} W^\pm$ (when the extra jets or
leptons in the deacy miss detection) and $W^\pm q^\prime W^\pm q^\prime$ (when
the two light quarks are misidentified as b-jets). Their
corresponding rates are found to be \cite{Like-sign,Barger,WW}
\begin{eqnarray}
& & \sigma ( t \bar{t} W^+ ) = 0.21  ~{\rm pb},
    ~~~\sigma (t \bar{t} W^- )= 0.1 ~{\rm pb} , \label{ttw}\\
&& \sigma (  W^+ q^\prime W^+  q^\prime ) = 0.5 ~{\rm pb},
    ~~~\sigma ( W^- q^\prime W^- q^\prime ) =0.23 ~{\rm pb} .
\end{eqnarray}
To effectively suppress the backgrounds and at the same time not
to hurt the signal too much, we search for the events with two
like-sign leptons plus exactly two jets in which at least one is
required to be a b-jet. Two-jets requirement can efficiently
suppress $t \bar{t} W$ backgound and one b-jet requirement can
eliminate most $WWqq$ background \cite{tt-background}. As a
result, the background can be suppressed by one order. The $S/B$
ratio can be further enhanced by imposing suitable kinematic cuts.
From the analysis of Ref. \cite{tt-background}, one may infer that
by assuming $60 \%$ b-tagging efficiency \footnote{ In Ref.
\cite{tt-background} a rather low b-tagging efficiency ($36 \% $)
was taken and thus more signal events were cut out.}, the
background can be reduced to $6$ events for $ 100 fb^{-1}$
integrated luminosity,  at the cost of a reduction of $86\%$ to
the signal. So, for an integrated luminosity $100$ fb$^{-1}$,
$\sigma( p p \to t t+X)$ larger than $10$ fb may be observable at
the LHC.


Note that in the TC2 theory there may exist other sources of FCNC
which may contribute to $ c c \to t t $. For example, the TC2 theroy
predicts a new gauge boson $Z^\prime $, which can also mediate flavor-changing
interactions \cite{tc2-Hill}. However, electroweak data constrained
$Z^\prime$ to be heavier than several TeV \cite{Electr-constrain},
and thus the effects of $Z^\prime$ are negligiblly small.

\subsection{$t t \bar{c}$ production at the LHC}

In the TC2 model the production $pp\to t t \bar{c}+X$ proceeds through
the patron process $ c g \to t t \bar{c}$, as shown in Fig.\ref{feynman} (b,c,d).
Like the process $ c c \to t t $, top-pion diagrams and top-Higgs diagrams
interfere destructively.
Since top-pion and top-Higgs may be produced on-shell in this process,
as shown in Fig.\ref{feynman} (b), we need to know their total widths.
The possible decay channels of top-pion (top-Higgs) are
\begin{eqnarray}
\pi_t^0 (h_t) \to  t\bar{t}, ~t\bar{c}, ~\bar{t} c, ~b\bar{b},
                  ~WW,   ~Z Z,      ~\gamma Z,  ~g g, ~\gamma \gamma
\label{decay}
\end{eqnarray}
For $m_t < m_{\pi_t^0, h_t} < 2 m_t$,
the process can be approximated as the direct production
of top-pion (top-Higgs) followed by their deacy to $t \bar{c}$.
Since the last five decay modes in Eq.(\ref{decay}) occur only at loop-level,
a moderate $K_{UR}^{tc}$ will make $t \bar{c}$ channel the dominant
decay mode of top-pion (top-Higgs). So in the region
$m_t < m_{\pi_t^0, h_t} < 2 m_t$, the cross section is
proportional to the square of $K_{UR}^{tc}$, less sensitive to
$K_{UR}^{tc}$ than in other parameter regions where the cross section
is proportional to $(K_{UR}^{tc} )^4 $.

\begin{figure}[hbt]
\begin{center}
\epsfig{file=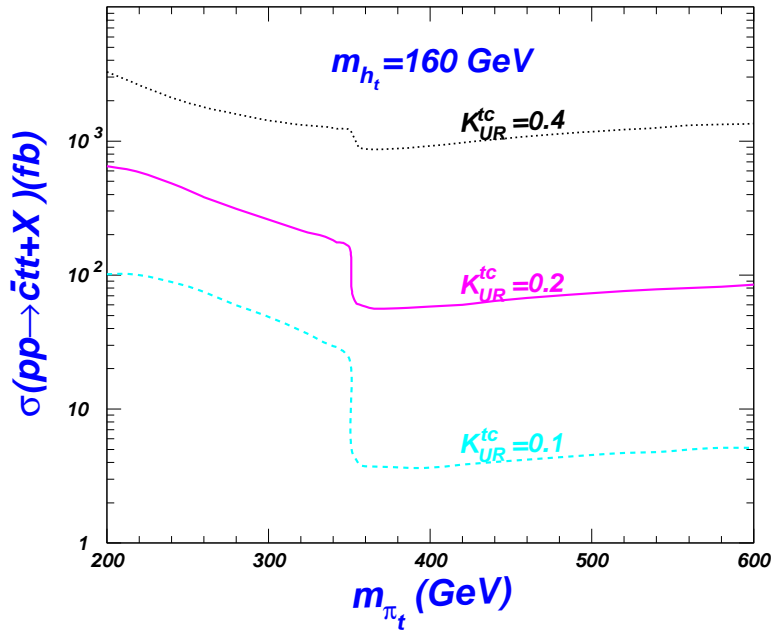,width=11cm}
\caption{Cross section of $ p p \to t t \bar{c} +X $ at the LHC as a function of
$m_{\pi_t}$ for various $K_{UR}^{tc}$.  }
\label{ttc}
\end{center}
\end{figure}

\begin{figure}[hbt]
\begin{center}
\epsfig{file=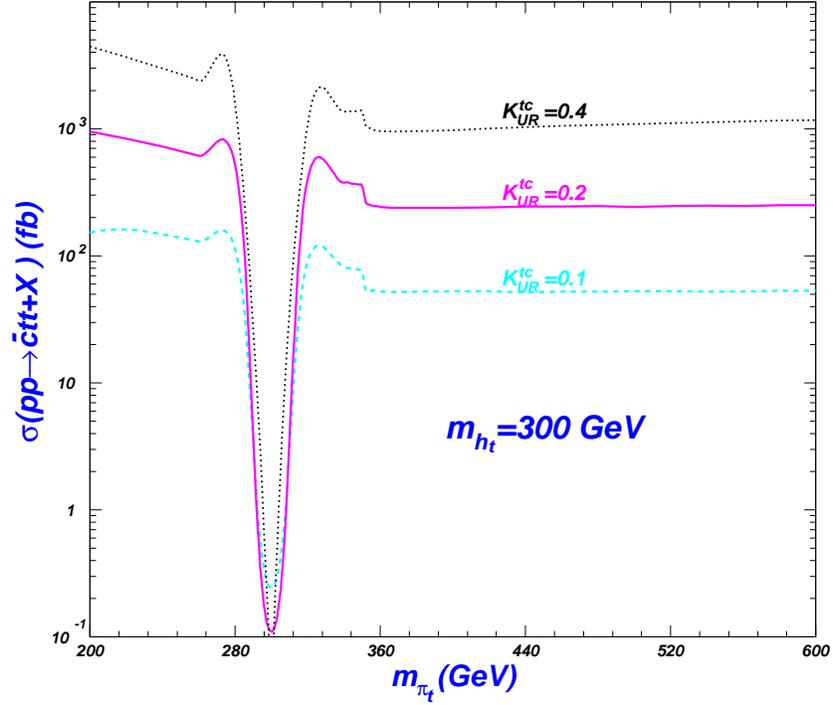,width=11.5cm}
\caption{Same as Fig.\ref{ttc}, but for $m_{h_t} =300$ GeV.}
\label{ttc1}
\end{center}
\end{figure}

\begin{figure}[hbt]
\begin{center}
\epsfig{file=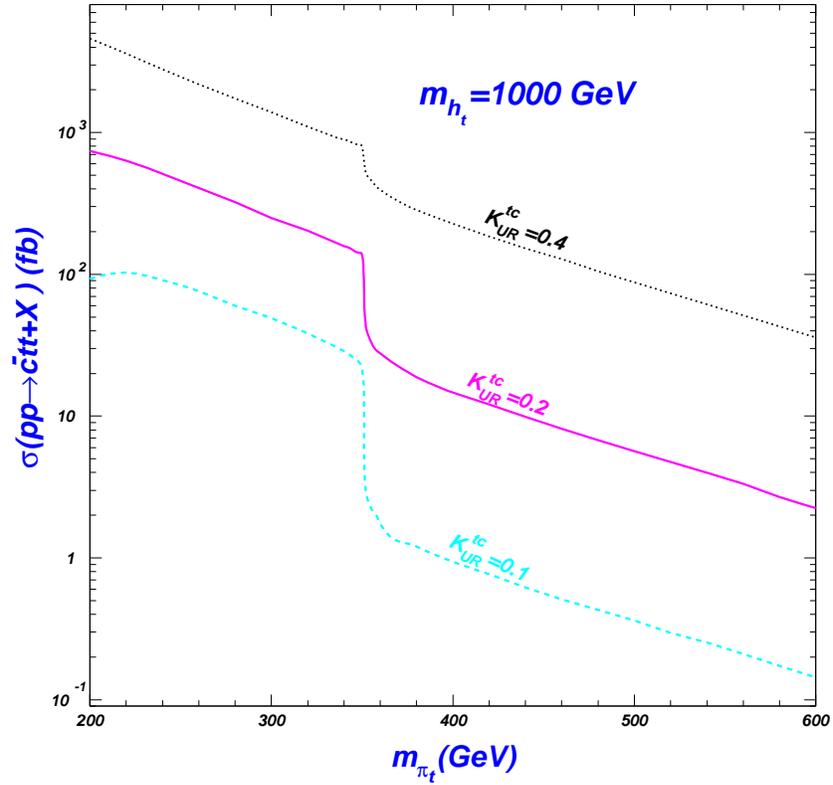,width=11.5cm}
\caption{Same as Fig.\ref{ttc}, but for fixed $m_{h_t}=1000 $ GeV.}
\label{ttc2}
\end{center}
\end{figure}

Figs.(\ref{ttc},\ref{ttc1},\ref{ttc2}) show the cross section of
$p p \to t t \bar{c}+X$ as a function of $m_{\pi_t^0}$ for various
$K_{UR}^{tc} $ and $m_{h_t}$. The charge conjugate production $pp
\to \bar t \bar t c + X$ is also taken into account. From these
figures, one can see that even for $K_{UR}^{tc} =0.1$, the cross
section can reach several tens fb in a sound parameter space, and,
depending on different parameter space, it may be larger or
smaller than the cross section of $ p p \to t t +X$. The sharp
drops of the cross section at $m_{\pi_t} \simeq 360$ GeV in these
figures reflect the suppression of $ Br(\pi_t^0 \to t \bar{c})$
due to the opening of decay channel $ \pi_t^0 \to t \bar{t} $.
Like Fig. \ref{pptt}, the dip of the cross section around
$m_{\pi_t}=300$ GeV in Fig. \ref{ttc1} is due to the cancellation
effects of top-pion and top-higgs diagrams.

The signature of $ p p \to t t \bar{c} +X$ is two like-sign dileptons,
two b-jets, one light quark jet  plus missing energy, i.e.,
$\ell^+ \ell^+ b b j + \E_slash$ ($\ell=e,\mu$).
The background is mainly from $pp\to W^+ t \bar{t} \to \ell^+
\ell^+ b b j_1 j_2 + \E_slash$ with either $j_1 $ or $j_2 $
missing detection.
If we require exactly three jets with at least one b-jet in the signal
events, then according to Fig.9 of Ref. \cite{tt-background}, about
$3/4$ of the background can be cut out so that $\sigma( W t \bar{t} ) < 100$ fb.
The ratio of signal to background can be further enhanced by
applying appropriate kinetic cuts \cite{tt-background}.
So the signal with a rate large than several tens of fb should be observable
at the LHC.

\begin{figure}[hbt]
\begin{center}
\epsfig{file=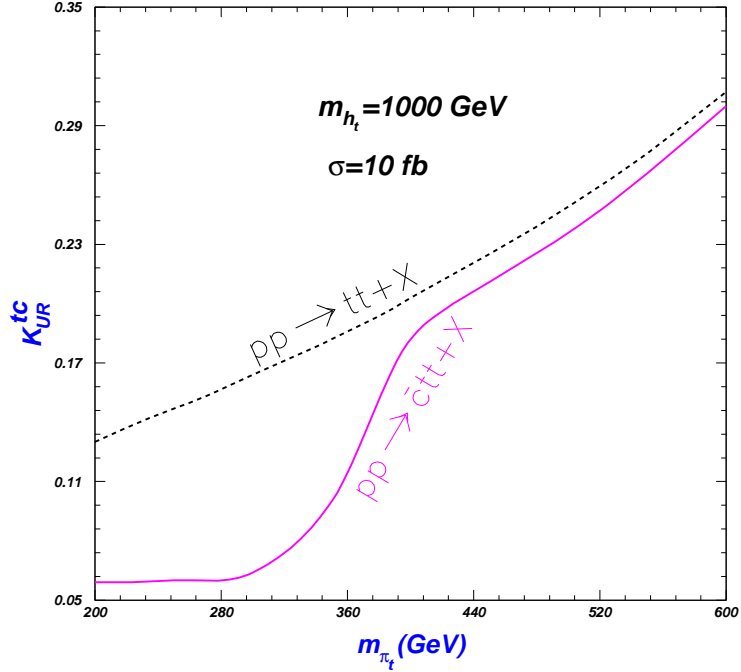,width=10cm}
\caption{The contour of the cross section for $ p p \to t t +X $
         and $p p \to t t \bar{c}+X $ at the LHC in $m_{\pi_t}$-$K_{UR}^{tc}$
         plane. }
\label{constant}
\end{center}
\end{figure}

A contour of the cross section in the $K_{UR}^{tc}$-$m_{\pi_t}$ plane
is plotted in Fig.\ref{constant}.  The region above each curve corresponds
to a cross section larger than  $10$ fb.
We see that in a large part of parameter space the cross section can exceed
$10$ fb for both processes.
We would like to make some comments on other like-sign top pair
production processes. First, we take a look at the production $ p
p \to t t \bar{c} q+X$, as shown in Fig.\ref{feynman} (e,f). At
first glance, this production may also have a sizable rate.
However, as found in the literature
\cite{Like-sign,Rainwater,Unitary}, due to the unitary constraint,
there exists severe cancellation between different diagrams so
that its rate is highly suppressed. We have calculated this
process and found that the cross section can maximally reach
several tens fb. But the background $t\bar t W^+$ in
Eq.(\ref{ttw}) is quite severe for this production. So it is not
as powerful as $tt$ and $t t \bar{c}$ productions in probing the
TC2 theory. The production $ p p \to t \bar{t} \pi_t^{0 \ast}
(h_t^\ast) \to t \bar{t} t \bar{c}$ \cite{Rainwater} can also lead
to like-sign top pairs in the final state at the LHC. But
analyzing its signal and background is quite complicated due to 
the multi particles in the final state. Particularly, if we 
require the two like-sign top quarks to decay semileptonically,
the reconstruction of this process may be quite difficult. 
We do not perform further analysis about these processes. 

Before ending this section, we want to point out that the
like-sign top pair productions may be quite unique in probing the
TC2 model at the LHC. To enhance the like-sign top pair production
rate to the accessible level at the LHC, the FCNC top quark
couplings $t \bar c \phi$ ( $\phi$ is any scalar field) or $t \bar
c V$ ($V=\gamma,Z,g$ or any new gauge boson) cannot be too small.
The TC2 model predict sizable tree-level $t \bar c \phi$ ($\phi$
is top-pion or top-higgs) coupling and thus may enhance the
like-sign top pair production rate to the accessible level at the
LHC. In many other popular extensions of the SM, there are no tree-level
top quark FCNC couplings and the couplings $t \bar c \phi$ and $t
\bar c V$ are induced at loop-level, which are usually too small to make
the like-sign top pair productions observable at the LHC. For
example, the top quark FCNC couplings are induced at loop-level in
the MSSM \cite{tcv1}. Although they can be much larger than in the
SM, we found that their contribution to the cross sections of $ p
p \to t t +X $ at the LHC is smaller than $10^{-4}$ fb. Note that
among the two-Higgs doublet models, the so-called type-III model
(2HDM-III) \cite{2HDM} allows tree-level FCNC $t \bar c \phi$
interactions. However, such couplings are related by the CKM
matrix with the flavor-changing charged-Higgs interactions, and
thus are severely constrained by low energy
data\cite{constrain-2hdm}. For the currently allowed parameter
space of 2HDM-III, we found that the cross section at the LHC can
maximally reach several tens fb for $ p p \to t t \bar{c}+X $ and
$10$ fb for $ p p \to t t+X $. Such rates just lie on the edge of
observation at the LHC. Therefore, the like-sign top pair
productions at the LHC cannot constrain the 2HDM-III efficiently.

\section{conclusion}
The TC2 theory predicts tree-level FCNC top quark Yukawa couplings with top-pions.
We examined various like-sign top pair productions induced by such FCNC
couplings at the LHC.
We found that the productions $pp\to t t +X$ and $ pp \to t t \bar{c} +X $
can reach several tens fb in a sound part of parameter space,
which may be observable due to the low backgrounds.
Since other popular new physics models like the MSSM cannot enhance these
rare productions to the observable level,  searching for these productions at the
LHC will serve as a powerful probe for the TC2 model.

\end{document}